\documentclass[pre,amsmath,amssymb,twocolumn]{revtex4-2}

\usepackage{graphicx}
\usepackage{dcolumn}
\usepackage{bm}

\usepackage[english]{babel}

\usepackage[utf8]{inputenc}

\usepackage{epsfig}
\usepackage{color}

\usepackage{stackengine}
\usepackage{multirow}

\newcommand{\del}{\partial}
\renewcommand{\(}{\left(} 
\renewcommand{\)}{\right)} 
\renewcommand{\[}{\left[} 
\renewcommand{\]}{\right]}

\newcommand{\bs}[1]{\boldsymbol{#1}}
\newcommand{\lr}[1]{\langle{#1}\rangle}

\renewcommand{\vec}{\mathbf}

\begin{document}

\title{Non-equilibrium cluster-cluster aggregation in the presence of anchoring sites} 

\author{Renaud Baillou}
\altaffiliation{current address: PMMH, ESPCI, 75005 Paris, France}
\affiliation{Institut de Biologie de l'ENS, Ecole Normale Supérieure, CNRS, Inserm, Université PSL, 46 rue d'Ulm, 75005 Paris, France}
\author{Jonas Ranft}
\email{jonas.ranft@ens.psl.eu}
\affiliation{Institut de Biologie de l'ENS, Ecole Normale Supérieure, CNRS, Inserm, Université PSL, 46 rue d'Ulm, 75005 Paris, France}

\date{\today}

\begin{abstract}

Non-equilibrium cluster-cluster aggregation of 
particles %
diffusing in or at the cell membrane %
has been hypothesized to lead to domains of finite size in different biological contexts such as lipid rafts, cell adhesion complexes, or postsynaptic domains in neurons. In this scenario, the desorption of particles balances a continuous flux to the membrane, imposing a cut-off on possible aggregate sizes and giving rise to a stationary size distribution. Here, we investigate the case of non-equilibrium cluster-cluster aggregation in two dimensions where diffusing particles and/or clusters remain fixed in space at specific anchoring sites, which should be particularly relevant for synapses but may also be present in other biological or physical systems. Using an effective mean-field description of the concentration field around anchored clusters, we derive an expression for their average size as a function of parameters such as the anchoring site density. We furthermore propose and solve appropriate rate equations that allow us to predict the size distributions of both diffusing and fixed clusters. We confirm our results with particle-based simulations, and discuss potential implications for biological and physical systems. 

\end{abstract}

\maketitle

\section{Introduction}

The aggregation of particles into larger structures is observed not only in 
inactive %
physical~\cite{smoluchowski1916drei,*smoluchowski1918versuch,einax2013colloquium} but also in biological systems, where it plays an important role e.g.~in the formation of larger protein assemblies in cells. Such protein clusters typically form in or at cellular membranes; examples range from the clustering of chemotactic receptors in bacteria~\cite{sourjik2004receptor,greenfield2009self} to the aggregation of cell-cell adhesion proteins in epithelia into dense clusters~\cite{quang2013principles}. More generally, aggregation processes have been hypothesized to govern the spatio-temporal organization of domains in cellular membranes e.g.~in the context of lipid raft formation~\cite{turner2005nonequilibrium,rautu2018size} or the maturation of the Golgi apparatus~\cite{vagne2020minimal}. Such instances of `healthy' aggregation are not to be confounded with the abnormal protein aggregation often linked to neurodegenerative diseases~\cite{tyedmers2010cellular}, which has also received theoretical attention~\cite{lenz2017geometrical,miangolarra2021two}. 

In the context of diffusion-limited growth, aggregates (domains) grow by the addition of individual particles or by fusion with other clusters (domain coalescence). Whereas the former typically occurs at the molecular scale in supersaturated solutions and is described by Lifshitz-Slyozov-Wagner theory~\cite{bray2002theory}, the latter case of cluster-cluster aggregation or domain coalescence is more relevant for (typically larger) strongly interacting particles with small or vanishing nucleation barrier and has been described by Smoluchowski coagulation equations ~\cite{smoluchowski1916drei,*smoluchowski1918versuch,quang2013principles,turner2005nonequilibrium,rautu2018size,miangolarra2021two}. 
Submonolayer molecular beam epitaxy, during which  monolayer islands form due to the diffusion and aggregation of deposited adatoms on the crystal surface, is another growth scenario that has been described using rate equations for the concentrations of islands of a given size~\cite{bales1997self,einax2013colloquium}. In the absence of a specific mechanism for growth arrest, eventually only one large component will remain and grow until the depletion of available particles. On the contrary, biological systems are generally out of equilibrium and can use energy to maintain a stationary state with finite aggregate or domain sizes. For example, arrested liquid-liquid phase separation (LLPS) has become a very active field of research since the discovery of its importance in the formation of P-granules in the \emph{C.~elegans} oocyte~\cite{brangwynne2009germline,hyman2014liquid,weber2019physics}. In the context of LLPS, 
liquid droplets of the minority phase are limited in size by %
chemical reactions that are driven out of equilibrium by externally imposed gradients. 
In the context of particle aggregation or domain coalescence in cellular membranes, the recycling of individual particles, clusters or membrane domains, i.e.~their removal from the surface, can serve to maintain a stationary distribution of sizes of diffusing clusters or domains despite continuous fusion and aggregation~\cite{quang2013principles,turner2005nonequilibrium,rautu2018size}.

Recently, a similar mechanism has been hypothesized to underly the formation of postsynaptic domains (PSDs) at inhibitory synapses~\cite{ranft2017aggregation} (but see~\cite{chapdelaine2021reciprocal}). In neurons, synaptic transmission depends on the concentration of transmembrane neurotransmitter receptors at PSDs due to transient interactions with scaffold proteins that form larger domains by homotypic interactions; both excitatory and inhibitory PSDs have been shown to be highly dynamic structures subject to continuous turnover of their constituent particles~\cite{choquet2013dynamic,maynard2023quantifying}, raising the question of how domains of a well-defined size may be maintained. In ref.~\cite{ranft2017aggregation}, the authors characterized the stationary size distributions that arise from the interplay of the (possibly size-dependent) diffusion of particles and clusters, aggregation, and the turnover of individual particles. Here, we will follow up on this earlier work by extending the analysis to include the existence of anchoring sites that can locally pin clusters at specific positions at the surface and hinder their diffusion. At synapses, transsynaptic adhesion complexes link the presynaptic bouton to the PSD~\cite{varoqueaux2006neuroligins} and can thus be expected to act as such anchoring sites in the postsynaptic membrane. Beyond the specific example of PSDs, our theory generalizes earlier approaches to describe two-dimensional diffusion-limited aggregation with turnover to the case where surface impurities or interactions with cytoplasmic or extracellular structures locally impedes the diffusion of protein clusters or membrane domains, with the aim to predict the stationary size distribution of fixed and diffusing domains as a function of parameters. 

This article is organized as follows. In Sec.~\ref{sec:diffusion_equation}, we first derive a self-consistent equation for the radius of the fixed domains based on the solution of the diffusion equation for particles outside the domains. While this approach provides insight into the scaling of the domain size with anchoring site density, it does not account for the distribution of sizes of domains. In Sec.~\ref{sec:rate_equations}, we then present the results for those stationary size distributions obtained by rate and master equations for diffusing and fixed domains, respectively. In Sec.~\ref{sec:simulations}, we finally compare our results to particle-based Brownian dynamics simulations and discuss the role of spatial order in the distribution of anchoring sites, before we conclude with a short summary discussion of our results in Sec.~\ref{sec:discussion}.

%

\section{Mean-field description based on the diffusion equation}
\label{sec:diffusion_equation}

\begin{figure}
\includegraphics[width=0.75\columnwidth]{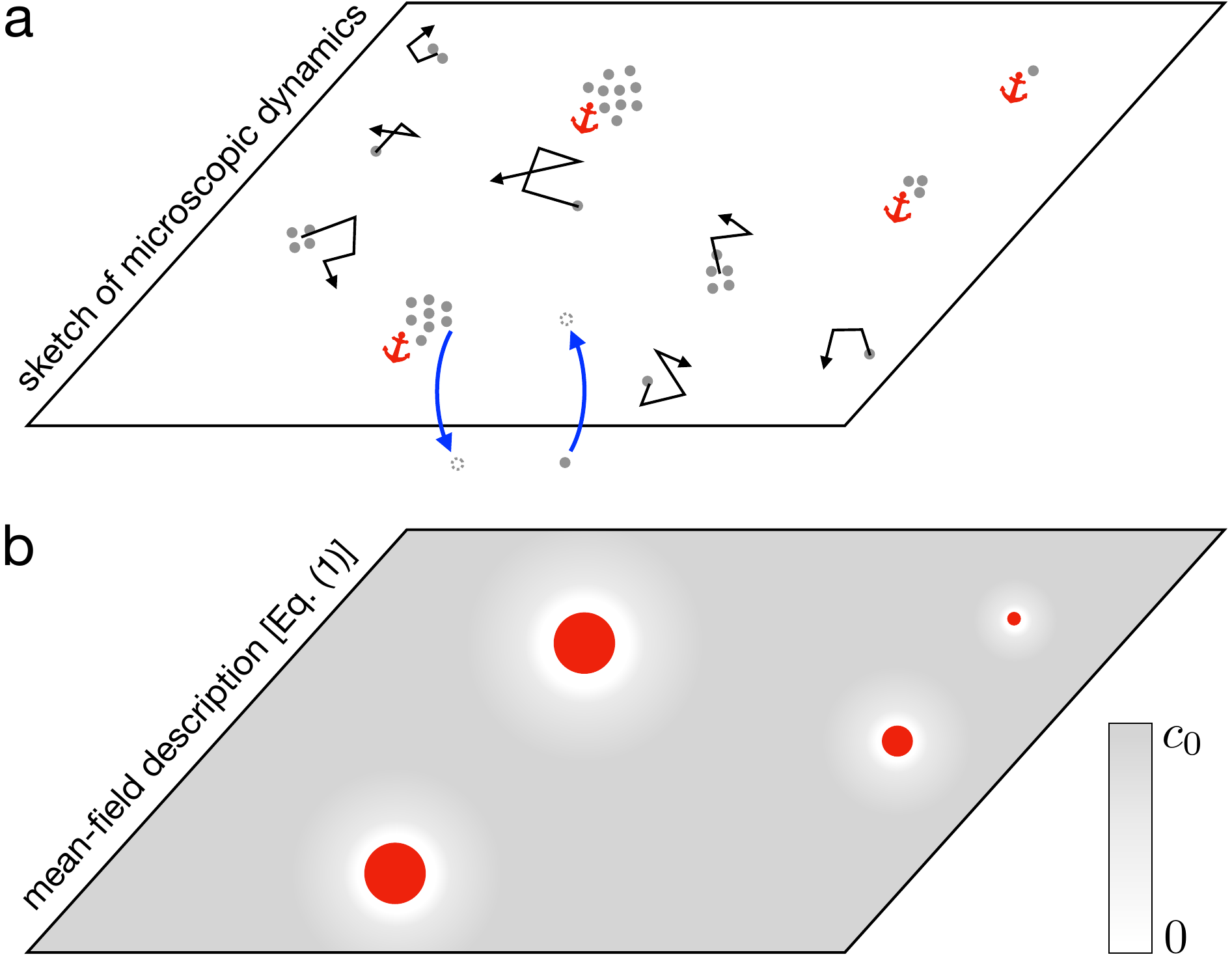}
\caption{
\label{fig:sketch}
Sketch of diffusion-aggregation dynamics with turnover in the presence of anchoring sites. (a) Particles and clusters that are not attached to anchoring sites can diffuse freely as indicated by the thin black arrows. Clusters at anchoring sites do not diffuse. Individual particles are subject to turnover (``monomer recycling scheme'') as indicated by the blue arrows. (b)  Corresponding mean-field picture with a smooth concentration field $c(\vec{r},t)$ (gray shading) outside of fixed domains (red).
}
\end{figure}

In a first approach, we describe the system as follows, see sketch of Fig.~\ref{fig:sketch}.
Diffusing particles outside of anchored domains are captured by a concentration field $c(\vec{r},t)$, $\vec{r} \in  \mathbb{R}^2$, which obeys the diffusion equation 
\begin{align}
\del_t c(\vec{r},t) = D \Delta c(\vec{r},t) - k c(\vec{r},t) + J
\end{align}
between fixed domains. Here, $D$ is the diffusion constant, $k$ is the recycling (or turnover) rate of individual particles, and $J$ is a flux that re-injects recycled particles into the system. Diffusion and turnover define a characteristic length scale $\lambda=\sqrt{D/k}$, and the recycling constants $k$ and $J$ define an equilibrium (average) concentration $c_0=J/k$. 

Upon encounter with a fixed domain, diffusing particles are adsorbed and added to that domain. Each domain $i$ thus acts as a local sink for $c(\vec{r})$ with a flux $B_i$ that depends on its radius $R_i$. In principle, this is described by absorbing boundary conditions for $c(\vec{r},t)$ at each domain boundary. However, for simplicity we consider point-like sinks located at anchoring sites $\vec{r_j}$, $j\neq i$, when describing the concentration field around a specific domain $i$, assuming that the typical extension of the domains is small compared to the distance between domains. In this limit, 
the concentration field obeys the equation
\begin{align}
\label{eq:DE_spec_sites}
\del_t c = D \Delta c - k c + J - \sum_{j\neq i} B_j \delta(\vec{r}-\vec{r_j}) \ .
\end{align}
The size $N_i$ of each domain $i$ expressed as the number of constituent particles obeys 
\begin{align}
\label{eq:dNidt}
\frac{d}{dt}N_i = B_i - kN_i \ ,
\end{align}
where $B_i$ is the flux of adsorbed particles and $kN_i$ is the particle loss due to turnover within the domain.

We can now try and coarse-grain Eq.~(\ref{eq:DE_spec_sites}) with respect to the (random) positions of the individual domains when all anchored domains are considered to be at steady state and of the same size $N$, i.e.~$B_i=kN_i=kN$. Stationarity furthermore implies a constant concentration field $c(\vec{r},t) = c(\vec{r})$. Introducing the density $n$ of anchoring sites and assuming radial symmetry around the specific domain that we consider, Eq.~(\ref{eq:DE_spec_sites}) then becomes 
\begin{align}
\label{eq:DE_coarsegrained}
0 = D \Delta c(r) - k(c(r)-c_0) - kNn \ ,
\end{align}
which needs to be solved with the absorbing boundary condition $c(R)=0$ at the domain edge $r=R$. One then readily obtains 
\begin{align}
\label{eq:c_of_r_solution}
c(r) = (c_0 - Nn)\(1-\frac{K_0(r/\lambda)}{K_0(R/\lambda)}\) \ ,
\end{align}
where $K_\alpha$ are the modified Bessel functions of the second kind.

\begin{figure}
\includegraphics[width=\columnwidth]{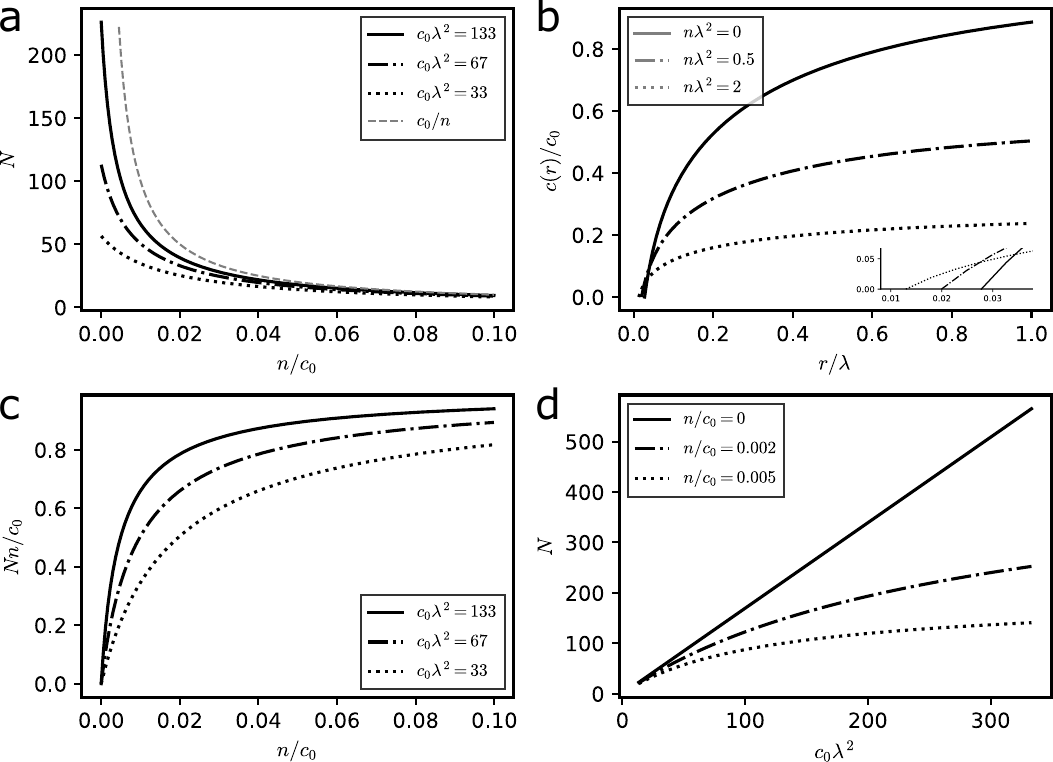}
\caption{
\label{fig:single_domain}
Mean-field theory for a single domain based on the diffusion equation. (a) Variation of the size of anchored domains with anchoring site density $n/c_0$. The asymptote $N=c_0/n$ when all mass is concentrated in fixed clusters is also shown, see text.  (b) Concentration field $c(r)$ outside of the anchored domains for different values of $n\lambda^2$. (c) Mass fraction of anchored domains as a function of anchoring site density $n/c_0$. (d) Size of anchored domains as a function of $c_0\lambda^2$ for fixed values of $n/c_0$. For all plots, $\rho/c_0=7000$. Cluster radii $R$ (and equivalently cluster sizes $N$) were obtained from numerically solving Eq.~\eqref{eq:R_implicit_simple}.
}
\end{figure}

The radius $R=\sqrt{N/(\pi\rho)}$ is linked to the domain size via the domain surface density $\rho$. The domain size $N$ itself follows self-consistently from the flux equilibrium condition $dN/dt=0$ (Eq.~(\ref{eq:dNidt})), namely 
\begin{align}
\label{eq:flux_eq_R}
kN = 2 \pi R D \del_r c(r)|_R \ . 
\end{align}
Here, the lateral influx is given by the diffusive particle flux $\vec{j}=-D\bs{\nabla} c(r)$ integrated along the domain circumference.
With Eq.~\eqref{eq:c_of_r_solution}, this condition leads to the following implicit equation for $R$:
\begin{align}
\label{eq:R_implicit_simple}
\frac{R}{\lambda} = 2 \[\frac{c_0}{\rho} - \pi \(\frac{R}{\lambda}\)^2 n\lambda^2 \] \frac{K_1(R/\lambda)}{K_0(R/\lambda)} \ .
\end{align}
The resulting domain size $N=\pi R^2\rho$ can thus be numerically determined as a function of the parameters of the system. For fixed
$c_0$, $\rho$, and $\lambda$, we show the predicted variation of $N$ with the density of anchoring sites $n$ in Fig.~\ref{fig:single_domain}a. The concentration field around anchored domains for different values of $n$ is shown in Fig.~\ref{fig:single_domain}b. 

When the density $n$ of anchoring sites increases, an increasing fraction of available particles remains trapped in the anchored clusters. We can obtain the mass fraction of anchored particles directly from the size $N$ of fixed domains via $Nn/c_0$, which we plot in Fig.~\ref{fig:single_domain}c as a function of $n/c_0$. Note that \emph{a priori} $N$ changes with $n$ in a non-trivial fashion (cf.~Eq.~\eqref{eq:R_implicit_simple}, Fig.~\ref{fig:single_domain}a); however, when almost all particles are trapped at anchoring sites, $N\approx c_0/n$ which we also plot on Fig.~\ref{fig:single_domain}a for comparison.

Finally, we can ask how the domain size $N$ changes with particle turnover rate $k$ for a fixed value of $n$, or more generally with $c_0$, $D$, and $k$ at a fixed relative value $n/c_0$. This is shown in Fig.~\ref{fig:single_domain}d. Note that while in the dilute limit ($n=0$) the domains can become arbitrarily large, their size can become at most $c_0/n$ (in the mean-field regime) when $n>0$, which corresponds to limiting case when all particles are trapped at anchoring sites.

So far, we have only considered aggregation at the fixed anchoring sites but completely ignored possible effects due to aggregation and cluster formation of diffusing particles outside of the anchored domains. Assuming a vanishing nucleation barrier for cluster formation of particles coming into contact, one however cannot exclude that particles diffuse as entire clusters in the free surface. In order to take into account the potentially non-negligible size of diffusing clusters, we introduce a typical radius $R_{\rm typ}$ and a typical diffusion constant $D_{\rm typ}$ of diffusing clusters and modify the above Eqs.~\eqref{eq:DE_coarsegrained} and \eqref{eq:flux_eq_R} accordingly, i.e., 
\begin{align}
0 &= D_{\rm typ} \Delta c(r) - k(c(r)-c_0) - kNn \ , \\
k R^2 \rho &= 2 R_{\rm eff} D_{\rm typ} \del_r c(r)|_{R_{\rm eff}} \ ,
\end{align}
where $R_{\rm eff} = R + R_{\rm typ}$ is the effective radius at which anchored and diffusing clusters fuse. With $\bar\lambda=\sqrt{D_{\rm typ}/k}$, the implicit equation for $R$ (Eq.~\eqref{eq:R_implicit_simple}) then becomes
\begin{align}
\label{eq:R_implicit_typ}
\frac{R^2}{\bar\lambda R_{\rm eff}} = 2 \[\frac{c_0}{\rho} - \pi 
\(\frac{R}{\bar\lambda}\)^2 n\bar\lambda^2 \] \frac{K_1(R_{\rm eff}/\bar\lambda)}{K_0(R_{\rm eff}/\bar\lambda)} \ .
\end{align}
To be able to make an educated guess about $R_{\rm typ}$ and $D_{\rm typ}$ if not to outright calculate them, we need a theory that allows to predict the size-distribution of diffusing clusters, which will be the subject of the next section.

\section{Cluster size distributions obtained from rate equations}
\label{sec:rate_equations}

In an earlier work, some of us have shown that in the dilute limit $n=0$, Smoluchowski rate equations capture extremely well the size distribution of diffusing clusters that one obtains in full particle-based simulations of Brownian dynamics with particle aggregation and turnover, in which a stationary size distribution is observed after a characteristic time $1/k$~\cite{ranft2017aggregation}. In the same spirit, we propose here rate equations that describe the evolution of the (spatially averaged) concentrations $c_m(t)$ of diffusing clusters of size $m$, where we introduce an additional term that accounts for the fusion with anchored domains. In particular, this approach allows us to take into account a possible size dependence of the diffusion constant of diffusing clusters, which we assume to be 
\begin{equation}
D_m = D_0 m^{-\sigma}
\end{equation}
in a scale-free way such that no additional length scale is introduced in the system. Once we have obtained the size distribution of diffusing clusters, we can use a master equation approach to calculate the size distribution of fixed clusters, providing us with a full description of the system that we set out to study.

In the presence of anchoring sites at a density $n$, the rate equations now read
\begin{multline}
\label{eq:rate_eq_with_n}
\frac{d}{dt} c_m = -kmc_m + k(m+1)c_{m+1} + J \delta_{m1} \\
- K \sum_j (D_j + D_m)c_j c_m + K \sum_{j<m} (D_j + D_{m-j})c_jc_{m-j}  \\
- K D_m c_m n \ ,
\end{multline}
where $K$ is a kinetic coefficient (see e.g.~ref.~\cite{hakim2020lifetime} for a discussion). In the dilute limit $n=0$ and for $\sigma=0$, an analytical solution for the stationary solution $c^*_m$ can be found~\cite{ranft2017aggregation}. For $n>0$ and/or $\sigma>0$ one can determine the $c^*_m$ numerically by integration of Eq.~\eqref{eq:rate_eq_with_n} until a stationary state is reached. It follows from the above equation that the normalized cluster concentrations $c^*_m/c_0$ only depend on three dimensionless parameters, 
\begin{align}
\label{eq:scaling_cm}
c^*_m = c_0 f_n(c_0\lambda'^2, \sigma, n/c_0)  \, ,
\end{align}
where we introduced the characteristic length scale $\lambda'=\sqrt{KD_0/k}$ reminiscent of $\lambda$ introduced in the previous section. 

While the stationary cluster size distribution depends on $n$ via the last term of Eq.~\eqref{eq:rate_eq_with_n}, it does not depend on the size $N$ of the fixed domains. To obtain $N$, we can proceed along two lines. Let us first invoke average particle number balance on the surface. The total concentration of particles $c_{\rm tot}$ is given by the weighted sum of concentrations $c_m$ plus the surface density of anchored particles which is given by $Nn$. At the same time, the balance of desorption and adsorption to the surface requires $kc_{\rm tot}=J$, or $c_{\rm tot}=c_0$. One thus obtains an average size for the anchored domains
\begin{align}
\label{eq:avN_from_mass_balance}
N = (c_0 - \sum_{m=0}^\infty m c^*_m)/n 
\end{align}
in the stationary state. Note that with Eq.~\eqref{eq:scaling_cm} it follows that $N$ only depends on the same three dimensionless parameters as $c^*_m/c_0$. 

Second, we can use a previously developed master equation approach to compute the distribution of sizes for fixed clusters that fluctuate in size due to the stochastic aggregation with diffusing clusters and recycling~\cite{hakim2020lifetime}. For the sake of completeness we repeat here the governing equation for probabilities $p_l(t)$ for a domain to be of size $l$,
\begin{multline}
\label{eq:ME}
\frac{d}{dt}p_l = -klp_l + k(l+1)p_{l+1} - p_l K \sum_j D_j c_j \\
+ K \sum_{j<l} D_{l-j} c_{l-j} p_j  
\end{multline}
for $l>0$, where the first two terms describe size changes due to the desorption of individual particles and the last two terms size changes due to fusion with diffusing aggregates. Note that strictly speaking the $p_l$ for $l\ge 1$ do not define a probability distribution, as $(d/dt)\sum_{l\ge1}p_l=-kp_1$ and $l=0$ is an absorbing state. However, one can define a quasi-stationary solution $\hat p_l$ such that $p_l(t)=\hat p_l e^{-\nu t}$ on long times, with $\nu \ll k$ and $\sum_{l\ge1}\hat p_l=1$. The $\hat p_l$ can be obtained in a semi-analytical way from a recursion relation~\cite{hakim2020lifetime} or via the relaxation of an initial $p_l(0)$ towards the (quasi-)stationary state by integrating Eq.~\eqref{eq:ME} numerically. Before solving for the $\hat p_l$, we can again make a scaling argument using Eq.~\eqref{eq:scaling_cm} and find they also depend only on $c_0\lambda'^2$, $\sigma$, and $n/c_0$. 


Assuming now that each anchoring site indeed harbors a fixed domain, the absolute concentrations of fixed domains of size $l$ are then simply given by 
\begin{align}
\hat c_l = n \hat p_l \ .
\end{align}
The total density of clusters of size $m$ follows as $c^{\rm tot}_m = c^*_m + \hat c_m$. Finally, following this approach the average size of anchored domains is found to be
\begin{align}
\label{eq:avN_from_pl}
N = \sum_l l \hat p_l \ ,
\end{align}
which happens to give basically the same result as Eq.~\eqref{eq:avN_from_mass_balance} when evaluated.

\begin{figure}
\includegraphics[width=\columnwidth]{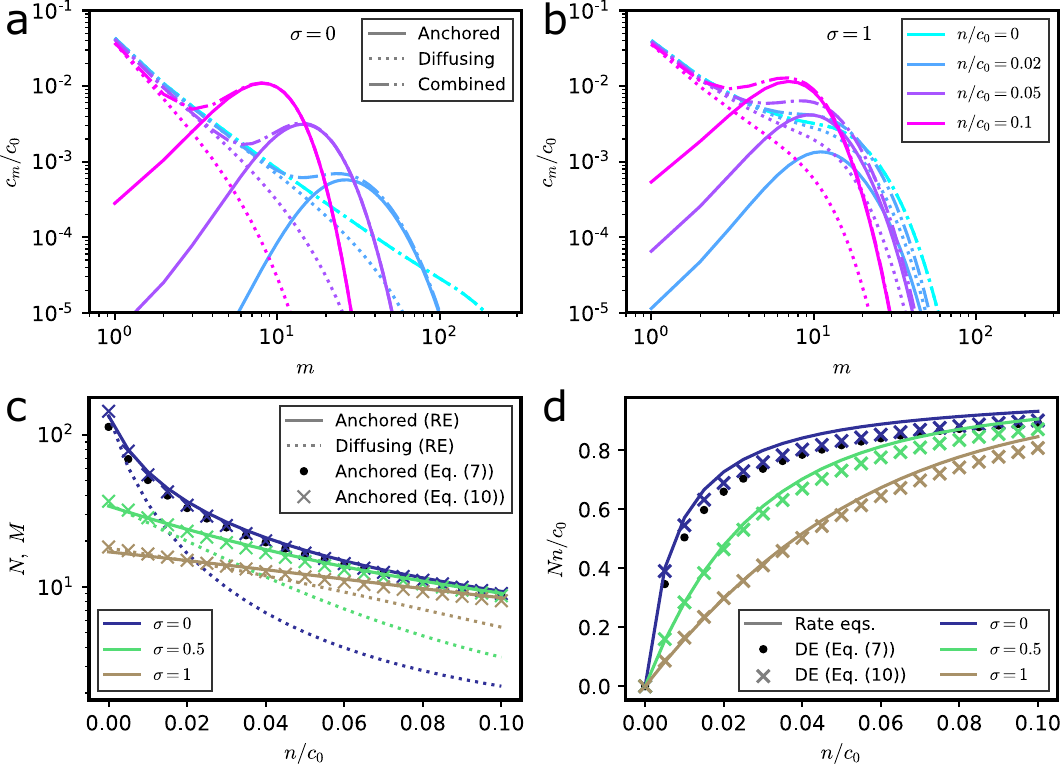}
\caption{
\label{fig:rate_equations}
Characterization of diffusing and anchored domains. (a,b) Size distributions for different values of $n/c_0$, for (a) $\sigma=0$ and (b) $\sigma=1$. Solid lines: $\hat c_m$, dotted lines: $c^*_m$, dashed-dotted lines: $c^{\rm tot}_m$; colors indicate the value of $n/c_0$, see legend in panel (b). 
(c) Average (typical) size of the anchored (diffusing) clusters as a function of $n/c_0$ for different values of $\sigma$. (d) Relative mass fraction of anchored domains as a function of $n/c_0$ for different values of $\sigma$. (c,d) The mean-field theory of Sec.~\ref{sec:diffusion_equation} is shown for comparison with dots (``naïve'' theory, Eq.~\eqref{eq:R_implicit_simple}) and crosses (``effective'' theory, Eq.~\eqref{eq:R_implicit_typ}). For the rate equations, $K=2$ was used.
}
\end{figure}

The results for the size distributions of diffusing and anchored clusters for different values of the anchoring site density $n$ and size-depedence exponent $\sigma$ of the diffusion constant are shown in Fig.~\ref{fig:rate_equations}a,b. In the dilute limit $n=0$, a strong dependence of the size distribution $c^*_m$ of diffusing clusters on $\sigma$ has been described~~\cite{ranft2017aggregation}. For $\sigma=0$, one can furthermore theoretically predict in this dilute limit that the size distribution follows a power law according to $c^*_m \sim m^{-1}$ until a typical size 
\begin{align}
\label{eq:Mdef}
M=\frac{\sum_m m^2c_m^*}{\sum_m mc_m^*}
\end{align}
that scales as $M^{\rm dil}_0\approx c_0\lambda'^2$. In the more general case of $\sigma>0$ in the dilute limit, the power law no longer holds, while the typical size of diffusing clusters follows the more general scaling $M^{\rm dil}_\sigma\approx (c_0\lambda'^2)^\frac{1}{1+\sigma}$. This picture changes in the presence of a finite $n>0$. Notably, the total concentration $c^{\rm tot}_m$ starts to deviate from a power law for $\sigma=0$ (Fig.~\ref{fig:rate_equations}a), and differences between the size distributions of diffusing clusters $c^*_m$ for different $\sigma$ become less pronounced with increasing $n$ (compare Fig.~\ref{fig:rate_equations}a,b). For increasing $n$, the size distributions of diffusing and anchored clusters overlap less and less, which manifests itself by an increasingly prominent trough in the total concentrations $c^{\rm tot}_m$ that separates both types of clusters. 

The average size $N$ of the fixed domains as a function of $n$ as predicted by the rate-equation approach is shown in Fig.~\ref{fig:rate_equations}c. We first observe that in line with the results of the previous section, $N$ decreases with the density of anchoring sites $n$. Moreover, the difference between the average size for different values of $\sigma$ does indeed decrease with $n$, as one can expect from the fact that it does only depend on the concentrations of diffusing clusters $c^*_n$, which become increasingly similar. With increasing $n$, the mass fraction $nN/c_0$ of anchored domains does increase despite decreasing $N$, as shown in Fig.~\ref{fig:rate_equations}d. Consequentially, as more and more mass is concentrated in the anchored domains, their average size must decrease as $N\approx c_0/n$ for large $n$, see Eq.~\eqref{eq:avN_from_mass_balance} (Fig.~\ref{fig:rate_equations}c).

Before we turn to particle-based simulations in the next section, we can already ask how well the ``effective'' theory of the previous section (Eq.~\eqref{eq:R_implicit_typ}) accounts for the average size $N$ of anchored domains when taking into account a typical radius $R_{\rm typ} =\lr{\lr{r_m}}$ and diffusion constant $D_{\rm typ}=\lr{\lr{D_m}}$ of diffusing clusters, where we defined $\lr{\lr{a_m}}={\sum_m m a_m c_m^*}/{\sum_m mc_m^*}$ to be the mass-weighted average over diffusing clusters and $r_m = \sqrt{m/(\pi\rho)}$. The comparison between the ``naïve'' theory (Eq.~\eqref{eq:R_implicit_simple}) with $D=D_0$, the ``effective'' theory with $R_{\rm typ}$ and $D_{\rm typ}$ informed from the rate-equation approach, and the result for $N$ from the rate-equation approach itself is also shown in Fig.~\ref{fig:rate_equations}. Perhaps not surprisingly, the corrections based on $R_{\rm typ}$ and $D_{\rm typ}$ almost entirely account for the discrepancy between the ``naïve'' theory and the result from the rate-equation approach.

\section{Particle-based simulations}
\label{sec:simulations}

\begin{figure}
\includegraphics[width=\columnwidth]{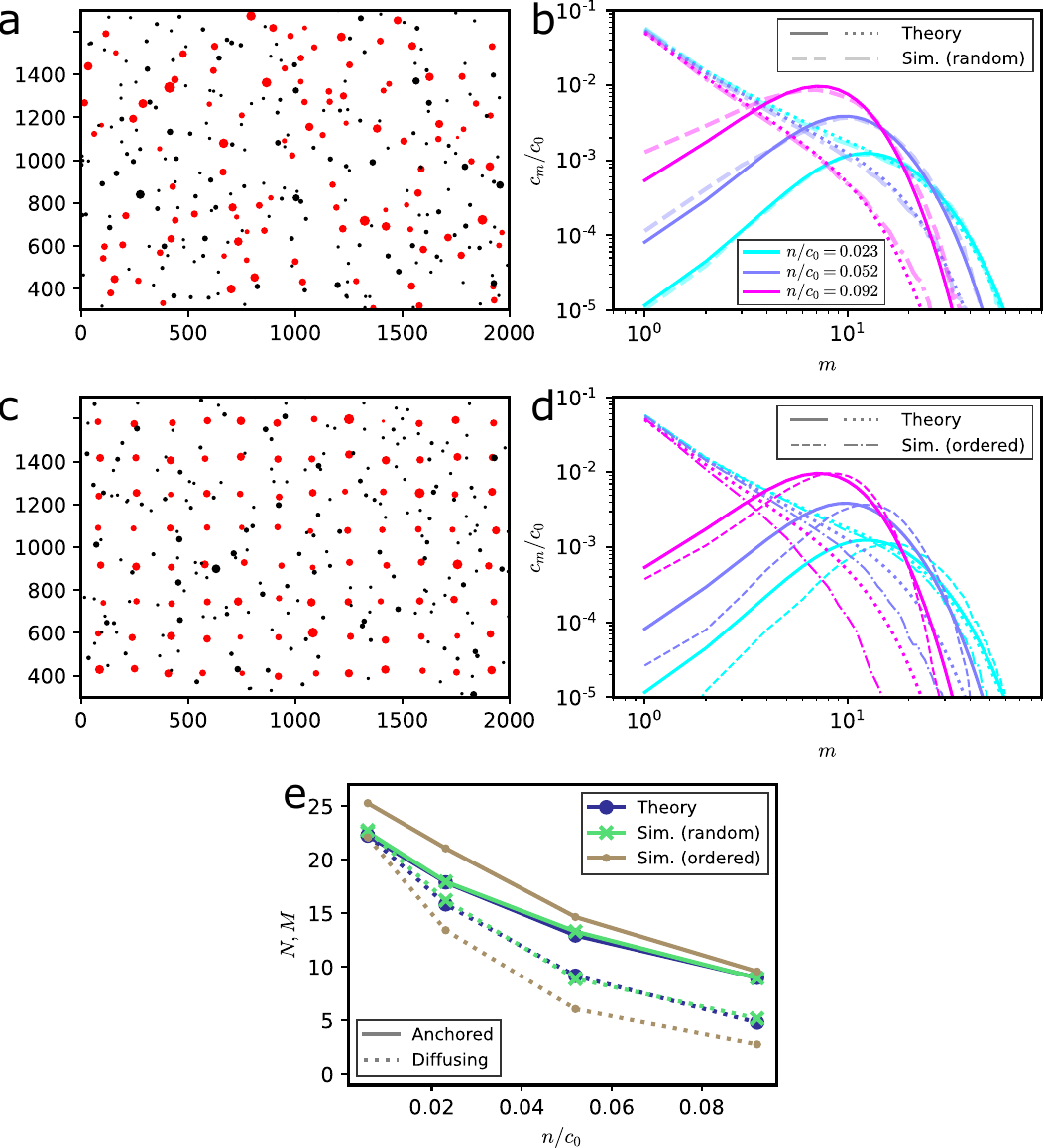}
\caption{
\label{fig:simulations}
Particle-based simulations. (a) Snapshot of a simulation  with 144 randomly distributed anchoring sites in a square box of side length $L=2000$. Lengths are in units of single particle diameter $a$, see refs.~\cite{ranft2017aggregation,hakim2020lifetime}; cluster radii are not to scale but increased five-fold for better visibility. Clusters at anchoring sites are colored in red.  (b) Size distributions of anchored and diffusing clusters for different values of $n/c_0$, comparing simulations with randomly distributed anchoring sites to the rate-equation-based theory. Note that the theory has no free parameter. (c,d) Same as (a,b) but for simulations with regularly distributed anchoring sites. (e) Average size $N$ of anchored clusters (solid lines) and typical size $M$ of diffusing clusters (dotted lines) as a function of $n/c_0$ as predicted by theory (blue, large dots) and obtained by simulations with randomly (green, crosses) or regularly (brown, small dots) distributed anchoring sites. For all simulations, parameters were identical to case II of ref.~\cite{hakim2020lifetime} ($c_0/\rho=9\cdot10^{-4}$, $c_0D_0/k=45$, $\sigma=0.5$), with a simulation box size $L=2000$ and for a duration $T=5\cdot10^7$. 
}
\end{figure}

To validate our theory, we implemented particle-based simulations following the model of refs.~\cite{ranft2017aggregation,hakim2020lifetime}. In brief, $c_0L^2$ particles were distributed on a square simulation box of lateral dimension $L$ with periodic boundary conditions. Upon diffusive encounter, particles fuse and form larger clusters, which can also diffuse. Here, we assumed that particles within clusters rearrange quickly and considered  disc-like clusters the radius of which is related to the number of constituent particles according to $R=\sqrt{N/(\pi \rho)}$,  where $\rho$ is the particle density within clusters, but note that we have shown in earlier work that the resulting cluster size statistics is almost identical in the absence of any particle rearrangement~\cite{ranft2017aggregation}. Individual particles (monomers or constituent particles of larger clusters) are removed from the surface with rate $k$ and randomly injected at the surface to keep the total particle number constant. In addition, $nL^2$ anchoring sites are distributed on the surface either randomly or forming a square lattice. While particles and clusters in principle diffuse with a size-dependent diffusion constant $D_m$, clusters that are in contact with these anchoring sites do not diffuse. Example snapshots of the simulations are shown in Fig.~\ref{fig:simulations}(a,c). We compare the simulation results for the observed size distributions of diffusing and anchored domains with the corresponding predictions from the rate-equation approach of Sec.~\ref{sec:rate_equations} in Fig.~\ref{fig:simulations}(b,d). (Note that the predictions are obtained without any free adjustable parameter as we chose the kinetic coefficient $K=1.81$ according to the calibration in the absence of anchoring sites of ref.~\cite{hakim2020lifetime}.) The comparison of the average (typical) size of fixed and diffusing clusters as a function of anchoring site density for both random and ordered cases with the theoretical predictions is shown in Fig.~~\ref{fig:simulations}(e). 

In the disordered case, the excellent agreement between the size distributions for specific parameter choices [Fig.~\ref{fig:simulations}(b)] and the dependence of average sizes on anchoring site density [Fig.~\ref{fig:simulations}(e)] allows us to conclude that our theory provides a quantitatively precise account of the aggregation dynamics and the ensuing non-equilibrium stationary state. 
As biological systems often display disorder, as can be seen for example in the random distribution of inhibitory PSDs on the dendritic membrane, our theory can thus be considered as quantitatively relevant beyond the general qualitative agreement between simulation and theory independent of spatial order. %
Interestingly, the agreement becomes less exact in the case of orderly distributed anchoring sites [Fig.~\ref{fig:simulations}(d,e)], in which case fixed clusters tend to be slightly larger then predicted by our theory, and diffusing clusters conversely slightly smaller. It seems to make intuitively sense that the disordered case is better described by the mean-field rate equations for the concentrations of domains of a given size, assuming that potential spatial correlations between neighboring clusters effectively self-average and cancel each other in this case. While it is not entirely clear why spatial correlations in the ordered case should give rise to an increased average size of the anchored domains, one may speculate that the regular tiling of space tends to suppress the growth of larger diffusing clusters and thus leads to the accumulation of a larger mass fraction of particles in fixed clusters. 

%

\section{Conclusion}
\label{sec:discussion}

To summarize, we developed a framework to predict the size distribution of diffusing and fixed domains that form by irreversible aggregation of particles combined with particle turnover 
in the presence of anchoring sites. In doing so, we followed three separate approaches that provided increasing insight into the  statistics of the resulting non-equilibrium stationary state. First, by considering the concentration field of diffusing particles around a stationary domain, we derived an implicit, self-consistent equation for the domain radius. The calculation predicts a decrease of the domain size with the density of anchoring sites and a concomitant increase in the fraction of particles at fixed domains (Sec.~\ref{sec:diffusion_equation}, Fig.~\ref{fig:single_domain}). We then combined Smoluchowski rate equations for the freely diffusing clusters with a Master equation for the stochastic size dynamics of fixed clusters to predict more precisely the shapes of the distribution of sizes of both types of cluster (Sec.~\ref{sec:rate_equations}, Fig.~\ref{fig:rate_equations}).  Notably, the results obtained using the diffusion-equation approach capture rather well the average size of fixed domains predicted by the rate- and Master equation approach [Fig.~\ref{fig:rate_equations}(e,f)]. Given that both approaches are quite different insofar as the first relies on the spatial gradient in the concentration field 
whereas the second discards spatial variations completely and focuses on the kinetics of aggregation instead, this is in and of itself indicative of the validity of our theory. Finally, we used particle-based simulations to assess the quantitative agreement between our mean-field theory and the aggregation dynamics in space (Sec.~\ref{sec:simulations}). Here, we found that a random distribution of anchoring sites leads to a non-equilibrium stationary state the size statistics of which are very well captured by our theory [Fig.~\ref{fig:simulations}(b,e)]. In the (biologically less relevant) case of 
ordered anchoring sites, the agreement becomes less precise, although our theory still captures 
qualitatively %
the dependence of the size distributions and average sizes on anchoring site density [Fig.~\ref{fig:simulations}(d,e)].

Our results indicate that the presence of anchoring sites modifies the cluster size statistics in several ways: (i) Diffusing and fixed clusters show distinct characteristic size distributions, 
the typical (average) sizes of %
which  
increasingly differ %
with increasing cluster site density; (ii) fixed clusters have a well-defined average size for arbitrary anchoring site density $n$ whereas (depending on the diffusion size-dependence exponent $\sigma$) diffusing clusters may show a power-law dependence until a cut-off size imposed by turnover at vanishing $n$, with large sizes becoming increasingly suppressed with increasing $n$. As a corollary, even if the diffusion constant  does not depend on cluster size ($\sigma=0$), the size distribution of diffusing and fixed clusters combined would show a deviation from a power-law scaling at finite anchoring site density $n$. Such a deviation is therefore not alone sufficient to indicate that diffusion is indeed size-dependent, as would be the case in the absence of any anchoring.

We note that the results we have obtained here for the case of a finite density of anchoring sites are somewhat similar to the finite-size effects discussed in the context of membrane nanodomains in vesicles~\cite{vagne2015sensing}, where the finite amount of aggregating particles leads to the separation of clusters into one ``giant'' cluster and small clusters when the system size becomes smaller or comparable to the typical cluster size in the infinite system. While the authors considered two distinct recycling schemes (monomer vs.~cluster recycling), they did not investigate the effects of a possible size-dependence of 
diffusion and used %
a simplified Master equation for the size evolution of the single large cluster; their study can thus be considered complementary to our work. In addition, our particle-based simulations reveal the subtle importance of spatial correlations in the resulting size statistics, which is a property of the extended system, i.e.~containing much more particles than the typical cluster size. It would be interesting to see whether analytical approaches beyond pure mean-field description may shed additional light onto these kinds of effects (see e.g.~\cite{bressloff2020active,bressloff2023two} for a discussion of the interaction between domains).

Another interesting consequence of the clear distinction between diffusing and fixed clusters is that the conceptual ``boundary'' between (actively suppressed) Ostwald ripening on the one hand~\cite{zwicker2015suppression,bressloff2020active} and domain or droplet coalescence on the other becomes somewhat blurred, as mass becomes increasingly concentrated in the larger fixed clusters which are replenished by diffusing small clusters. 

Here, we restricted ourselves to the statistics of the non-equilibrium stationary state that can arise in systems where aggregation is combined with 
turnover~\cite{turner2005nonequilibrium,vagne2015sensing}. %
Recently, the dynamics of the cluster size distribution for size-dependent recycling has been analytically studied using a continuum version of Smoluchowski coagulation equations, including results for the transient dynamics of the average cluster size for given initial conditions~\cite{rautu2018size}. It would be worthwhile to see whether these approaches may be extended to the case with size-dependent diffusion coefficient and a finite density of anchoring sites. Given the quantitative agreement between our stationary solution and the simulations, we would expect that the time evolution of the cluster size distribution from non-stationary initial conditions would equally well be captured using the time-dependent (numerical) solutions of the rate equations and Master equation, although admittedly this does not amount to an analytical theory.

This work has been inspired by the observation that inhibitory neurotransmitter receptors and scaffold proteins can spontaneously organize into surface domains of a well-defined size even in non-neuronal cells~\cite{calamai2009gephyrin,haselwandter2011formation}, which lead to the hypothesis that these domains form according to the aggregation process with turnover described here~\cite{ranft2017aggregation}. 
A %
recent study assessed whether the proposed non-equilibrium recycling scheme could explain the observed receptor and scaffold kinetics at synapses upon the suppression of lateral receptor fluxes at the extrasynaptic membrane, and concluded that direct recruitment of scaffold proteins to synaptic sites cannot be neglected~\cite{chapdelaine2021reciprocal}. Given the ubiquity of protein diffusion and aggregation in membranes, and the potential sites for protein anchoring at the submembrane structures such as the cytoskeleton, transmembrane or extracellular structures, we are confident that the results presented here remain nevertheless relevant for a large number of biological systems.

{\bf Acknowledgments.} The authors would like to thank Vincent Hakim and Benjamin Lindner for comments on an earlier version of this manuscript.

\bibliographystyle{apsrev}
\bibliography{refs.bib}

\end{document}